\documentclass[twocolumn]{aastex631}
\usepackage[utf8]{inputenc}
\usepackage{t1enc}
\usepackage{amssymb}
\usepackage{graphicx}
\usepackage{amsmath}
\usepackage{natbib}
\usepackage{booktabs}
\usepackage{rotating}
\usepackage{hyperref}
\usepackage[hyphens]{url}
\usepackage{silence}
\usepackage{etoolbox}
\usepackage{supertabular}
\usepackage{soul}
\usepackage{txfonts}
\usepackage{xcolor}
\usepackage[flushleft]{threeparttable}
\newcommand{\gcc}{g\,cm$^{-3}$}

\begin{document} 
\sloppy

\title{Three fast-spinning medium-sized Hilda asteroids uncovered by TESS}

\submitjournal{\apjl}

\correspondingauthor{N\'ora Tak\'acs} \email{takacs.nora@csfk.org}

\author[0009-0008-2021-1098]{N\'ora Tak\'acs}
\affiliation{Konkoly Observatory, HUN-REN Research Centre for Astronomy and Earth Sciences, Konkoly Thege 15-17, H-1121~Budapest, Hungary}
\affiliation{CSFK, MTA Centre of Excellence, Budapest, Konkoly Thege 15-17, H-1121, Hungary}
\affiliation{ELTE E\"otv\"os Lor\'and University, Institute of Physics and Astronomy, Budapest, Hungary}

\author[0000-0002-8722-6875]{Csaba Kiss}
\affiliation{Konkoly Observatory, HUN-REN Research Centre for Astronomy and Earth Sciences, Konkoly Thege 15-17, H-1121~Budapest, Hungary}
\affiliation{CSFK, MTA Centre of Excellence, Budapest, Konkoly Thege 15-17, H-1121, Hungary}
\affiliation{ELTE E\"otv\"os Lor\'and University, Institute of Physics and Astronomy, Budapest, Hungary}

\author[0000-0002-1698-605X]{R\'obert Szak\'ats}
\affiliation{Konkoly Observatory, HUN-REN Research Centre for Astronomy and Earth Sciences, Konkoly Thege 15-17, H-1121~Budapest, Hungary}
\affiliation{CSFK, MTA Centre of Excellence, Budapest, Konkoly Thege 15-17, H-1121, Hungary}

\author[0000-0002-5481-3352]{Emese Plachy}
\affiliation{Konkoly Observatory, HUN-REN Research Centre for Astronomy and Earth Sciences, Konkoly Thege 15-17, H-1121~Budapest, Hungary}
\affiliation{CSFK, MTA Centre of Excellence, Budapest, Konkoly Thege 15-17, H-1121, Hungary}

\author[0000-0002-1663-0707]{Csilla E. Kalup}
\affiliation{Konkoly Observatory, HUN-REN Research Centre for Astronomy and Earth Sciences, Konkoly Thege 15-17, H-1121~Budapest, Hungary}
\affiliation{CSFK, MTA Centre of Excellence, Budapest, Konkoly Thege 15-17, H-1121, Hungary}
\affiliation{ELTE E\"otv\"os Lor\'and University, Institute of Physics and Astronomy, Budapest, Hungary}

\author[0000-0002-0606-7930]{Gyula M. Szab\'o}
\affiliation{Konkoly Observatory, HUN-REN Research Centre for Astronomy and Earth Sciences, Konkoly Thege 15-17, H-1121~Budapest, Hungary}
\affiliation{ELTE Eötvös Loránd University, Gothard Astrophysical Observatory, Szent Imre h. u. 112, 9700, Szombathely, Hungary}
\affiliation{MTA-ELTE Exoplanet Research Group, Szent Imre h. u. 112, 9700, Szombathely, Hungary}

\author[0000-0002-8159-1599]{L\'aszl\'o Moln\'ar}
\affiliation{Konkoly Observatory, HUN-REN Research Centre for Astronomy and Earth Sciences, Konkoly Thege 15-17, H-1121~Budapest, Hungary}
\affiliation{CSFK, MTA Centre of Excellence, Budapest, Konkoly Thege 15-17, H-1121, Hungary}
\affiliation{ELTE E\"otv\"os Lor\'and University, Institute of Physics and Astronomy, Budapest, Hungary}

\author[0000-0002-1698-605X]{Kriszti\'an S\'arneczky}
\affiliation{Konkoly Observatory, HUN-REN Research Centre for Astronomy and Earth Sciences, Konkoly Thege 15-17, H-1121~Budapest, Hungary}
\affiliation{CSFK, MTA Centre of Excellence, Budapest, Konkoly Thege 15-17, H-1121, Hungary}

\author[0000-0002-3258-1909]{R\'obert Szab\'o}
\affiliation{Konkoly Observatory, HUN-REN Research Centre for Astronomy and Earth Sciences, Konkoly Thege 15-17, H-1121~Budapest, Hungary}
\affiliation{CSFK, MTA Centre of Excellence, Budapest, Konkoly Thege 15-17, H-1121, Hungary}
\affiliation{ELTE E\"otv\"os Lor\'and University, Institute of Physics and Astronomy, Budapest, Hungary}

\author[0000-0002-8585-4544]{Attila B\'odi}
\affiliation{Department of Astrophysical Sciences, Princeton University, 4 Ivy Lane, Princeton, NJ 08544, USA}
\affiliation{Konkoly Observatory, HUN-REN Research Centre for Astronomy and Earth Sciences, Konkoly Thege 15-17, H-1121~Budapest, Hungary}
\affiliation{CSFK, MTA Centre of Excellence, Budapest, Konkoly Thege 15-17, H-1121, Hungary}

\author[0000-0001-5449-2467]{Andr\'as P\'al}
\affiliation{Konkoly Observatory, HUN-REN Research Centre for Astronomy and Earth Sciences, Konkoly Thege 15-17, H-1121~Budapest, Hungary}
\affiliation{CSFK, MTA Centre of Excellence, Budapest, Konkoly Thege 15-17, H-1121, Hungary}
\affiliation{ELTE E\"otv\"os Lor\'and University, Institute of Physics and Astronomy, Budapest, Hungary}

\begin{abstract}
Hilda asteroids, which orbit in a 3:2 resonance with Jupiter, serve as key indicators of dynamical processes in the early solar system. Their spin rates, an important probe of these mechanisms, can constrain their density and collisional evolution, offering valuable insights into their origin.
In this paper, we report on the identification of three fast-rotating Hilda asteroids with spin periods in the 3.2--3.7\,h range using data from the Transiting Exoplanet Survey Satellite. These rotation periods are significantly shorter than the previous $\sim$5.0\,h shortest rotation periods obtained from ground-based observations in the $\sim$10\,km size range, and are comparable with the $\sim$3.0\,h breakup limit of Hildas a few km in size, derived from the FOSSIL survey. 
These fast-rotating asteroids require either considerable cohesion (in the order of a few kPa), or densities $\rho$\,$\gtrsim$1.5\,\gcc, in contrast to the typically assumed $\rho$\,$\lesssim$1\,\gcc, to prevent rotational break-up. 
C-type asteroids, which are common in the outer main belt, have densities of $\rho$\,$\approx$1.5\,\gcc\ and are known to comprise a small but notable fraction of Hildas. 
The observed occurrence rate of the $\leq$4\,h rotation periods may be explained by the 10-15\% fraction of C-type asteroids, likely mixed into these populations from the outer main belt during giant planet dynamical interactions in the early solar system.

\end{abstract}

   \keywords{Jovian trojans --
                TESS --
                Solar System --
                asteroids
               }

\section{Introduction}

Hilda asteroids are a resonant population of small bodies in the Solar System, orbiting the Sun in a 3:2 mean motion resonance with Jupiter. These objects spend most of their time far from Jupiter on their eccentric orbits, forming a stable triangular structure that shifts along their paths over time. Unlike dynamically unstable populations in nearby mean-motion resonances, Hildas are long-lived, with many members likely originating from the primordial outer Solar System \citep[e.g.,][]{2008MNRAS.390..715B, 2015AJ....150..186R}. Their stability within the resonance, combined with their compositional similarity to Jovian trojans, suggests a connection between these populations and the larger-scale dynamical processes that shaped the Solar System \citep[][]{2017AJ....154..104W, 2019P&SS..169...78Y,2025arXiv250304403V}.

The physical and rotational properties of asteroids provide insight into their internal structure and collisional history. Hildas exhibit a bimodal spectral distribution, with most classified as either D- or P-type asteroids, similar to Jovian trojans \citep{1995A&A...302..907D, 1997A&A...323..606D, 2008Icar..193..567G, 2017AJ....153..145W, 2017AJ....154..104W}. The existence of collisional families, such as those associated with (153) Hilda and (1911) Schubart, further suggests that disruptive impacts have played a role in shaping this population \citep{2008MNRAS.390..715B}.

According to the Kepler K2 observations,
Hilda asteroids show a significantly lower number of fast rotators (f\,$\geq$7\,c/d) and an overpopulation of slow rotators (f\,$\leq$\,24\,c/d) with respect to main belt asteroids \citep{Szabo2020}. Fast rotators are important because they set the break-up limit, the fastest spin rate that a rotating asteroid can survive without falling apart \citep{Pravec2000}. Assuming a rubble pile structure and that gravity is the only force that keeps the asteroid together, this break-up limit defines a critical density, an upper limit on the possible densities within a specific population. In the main belt, the $\sim$2.2\,h breakup limit corresponds to a critical density of $\sim$2.2\,\gcc\ for asteroids larger than a few kilometers in size. Among Hildas, previous ground-based observations set a break-up limit of $\sim$5\,h, as obtained from the Light Curve Database \citep[LCDB][using the update of October 1, 2023]{Warner2009}, similar to the traditional breakup limit of Jovian trojans, which are believed to have a similar origin to the Hildas \citep[see e.g.][for a summary]{Mottola2024}. 
In the 1--2\,km size range, the FOSSIL survey \citep{Chang2022} found a number of Hildas that rotate with periods of $\sim$3--4\,h, setting an updated $\sim$3\,h breakup limit for small Hilda asteroids. Recently, \cite{2025A&A...694L..17K} identified the fastest rotating Jovian trojans known currently, with the fastest rotation period being P, =, 2.95\,h for (13383).
This fast rotation can either be explained by a strengthless rubble pile structure and a relatively high density of $\sim$1.6\,\gcc, or by a considerable cohesion, in the order of a few kPa.  
In this paper, we report on the identification of three fast-rotating Hilda asteroids (42237), (91273) and (237321) using data mainly from the Transiting Exoplanet Survey Satellite \citep[TESS,][]{Ricker2015}. 
Supplemented by two fast-rotating Hildas from the Kepler/K2 mission \citep[(207644) and (208290),][]{Szabo2020} these targets set the fastest rotating Hilda sample in the $\sim$10\,km size range. We describe our observations and data reduction steps in Sect.~\ref{sect:obs}, present our results in Sect.~\ref{sect:results}, and discuss the significance of these newly identified fast rotation rates in Section~\ref{sect:discussion}.

   \begin{figure*}
   \centering
    \includegraphics[width=\textwidth]{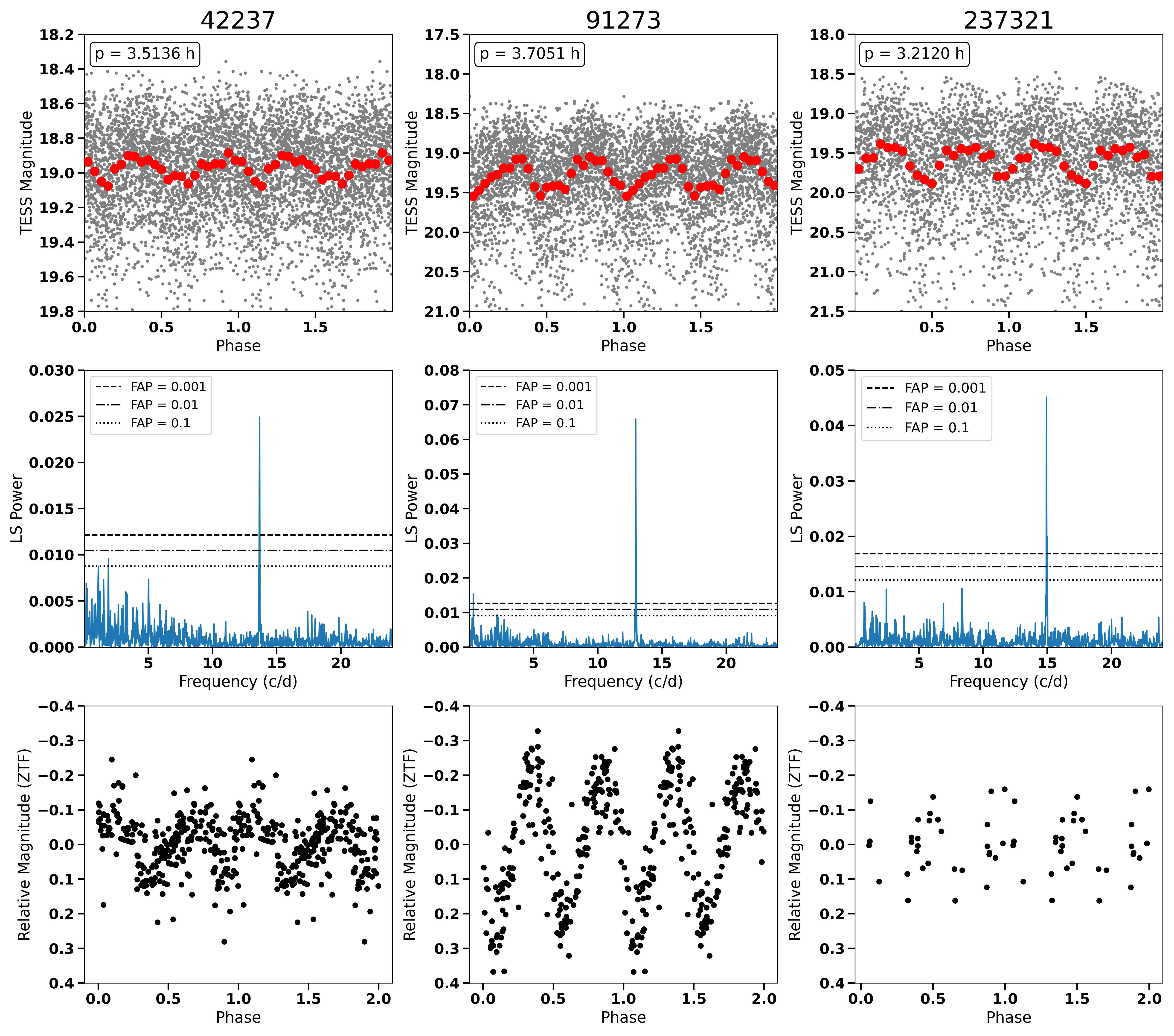}
   \caption{Folded light curves and rotational frequencies for the three fast-rotating Hilda asteroids. Top panel: light curves are folded with the obtained rotation periods, also listed in Table~\ref{table:targets}. The grey points, representing the observations, are binned with 24, 26 and 22 bins for asteroids (42237), (91273) and (237321), respectively. Middle panel: power spectra with the false alarm probabilities (FAP) marked with dashed lines. Bottom panel: light curves obtained from the ZTF survey and folded with periods, indicated on the figures.} \label{fig:lcs}
  
    \end{figure*}

\section{Observations and data reduction}
\label{sect:obs}

Photometric data for fast-rotating Hilda asteroids (42237), (91273), and (237321) were obtained from the Transiting Exoplanet Survey Satellite (TESS) observations across multiple sectors as part of our TESS asteroid light curve survey, which includes $\sim$400 Hilda asteroids. 
Specifically, asteroid (42237) was measured in sectors 42, 43 and 44, (91273) in sectors 45, 46, and (237321) in sector 42. We also analysed the TESS data of the two K2 fast-rotating Hildas (207644) and (208290), in TESS sectors 42 and 43, respectively.

Data reduction follows an approach similar to the previous TESS-based analysis, first developed by \cite{Pal2020}, along with refinements applied to the pipeline since then, to optimise the extraction of asteroid light curves as discussed in \citet{2025A&A...694L..17K} and \citet{2025arXiv250313332T}. The processing pipeline is based on differential photometry using elongated apertures to maximise the signal-to-noise ratio for asteroids of varying brightness. Background contamination from variable stars and other sources was mitigated through adaptive noise propagation techniques, and automated masking was applied to remove data points affected by mutual asteroid crossings. Photometric extraction was performed using the FITSH package \citep{Pal2012}.
A detailed description of the data reduction modified with respect to the steps in \citet{Pal2020} can be found in \citet{2025arXiv250313332T}.

Following initial extraction, light curves were subjected to post-processing to remove residual systematics. The elimination of outliers was carried out using an iterative sigma-clipping method combined with the LOWESS algorithm \citep{Cleveland}, and instrumental trends and phase angle variations were corrected using low-order polynomial fitting \citep{Pal2020, Vavilov2025}. The final rotation periods were determined using Lomb-Scargle periodograms. 
To check for potential multiplicities of the primary frequencies identified, we also used the technique of higher order Fourier series \citep{Harris1989,Vavilov2025}, assuming Fourier orders up to i\,=\,8. 

Due to the large TESS pixel size of 21\arcsec\, the photometry is affected by changes in the position of the photometric aperture relative to the pixels. This may lead to characteristic frequencies in the residual spectrum related to the X and Y motions of the target in the reference frame of the CCD. To check the impact of this effect on our targets, we calculated the frequency spectra of the measured brightness values against the X and Y pixel positions, the same way as it was done by \citet{Makemake-2024} and ruled out any contamination from these effects.

\section{Results \label{sect:results}}

The results of the period search are presented in Table~\ref{table:targets}, along with complementary data (including absolute magnitude, albedo, diameter, and redness, when available). Figure~\ref{fig:lcs} also shows the phase-folded light curves and the rotational frequencies obtained. All three targets show a double-peaked light curve, and the higher-order Fourier fits show that the light curves cannot be better described by i\,$\geq$3 multiples of the base period. The rotation period of (91273) was also obtained by \citet{Durech2023} from Gaia photometry and was found to be identical to our TESS period, P\,=\,3.7051\,h. In Fig.~\ref{fig:lcs}, the significance levels of each power spectrum are defined by false alarm levels corresponding to probabilities of 0.1, 0.01 and 0.001 \citep{Scargle1982}.
In Table~\ref{table:targets}, we also include the amplitudes of the two K2 targets obtained from the TESS measurements. The periods of these faint targets could not independently be derived from the TESS data, and we used the K2 periods to fold the light curves. The amplitudes obtained are just upper limits in these TESS sectors (0.27$\pm$0.19, and 0.13$\pm$0.19\,mag for (207644) and (208290), respectively), which may be significantly different from the K2 amplitudes due to the different observation geometries and spin axis aspect angles.

\renewcommand{\arraystretch}{1.15}
\begin{table*}[!ht]
    \begin{center}
        
         \begin{tabular}[t]{c|ccccc}
        \hline
        \hline
          \textbf{ID}                        &   \textbf{(42237)}   &  \textbf{(91273)} &  \textbf{(237321)} & \textbf{(207644)}  & \textbf{(208290)} \\ 
          \textbf{SCC}                       &        S44C1C1       &      S45C3C1      &      S42C1C1/2     &    K2     &      K2     \\
          \textbf{P (h)}                     &  3.5136$\pm$0.0005    &  3.7051$\pm$0.0005 &  3.2120$\pm$0.0004  &   2.689   &    3.643    \\
          $\mathbf{\Delta m}$ \textbf{(mag)} &   0.21$\pm$0.02      &   0.53$\pm$0.02   &    0.53$\pm$0.003  &   0.23    &    0.36     \\
          $\mathbf{H_V} $ \textbf{(mag)}     &       12.30          &       13.45       &        14.22       &  14.74    &   14.58     \\
          \textbf{p$_V$ (km)}                &  0.064$\pm$0.008$^*$ &      0.061        &        0.061       &   0.061   &     0.061   \\
          \textbf{D (km)}                    &  18.14$\pm$0.25$^*$  &      11.0         &         7.7        &   6.1     &     6.5     \\
          \textbf{Redness}                   &         {LR}         &        -          &          -         &    LR     &      -      \\
          \textbf{$\overline{(g-r)}$}           &   0.60$\pm$0.07      &   0.52$\pm$0.07   &   0.55$\pm$0.16    & 0.59$\pm$0.09 & 0.32$\pm$0.04 \\
        \hline
         \end{tabular}
           \end{center}
        \caption{Main results of the asteroid light curve period determination. The columns of the table are: ID: asteroid number; SCC: TESS Sector, Camera and Chip identifier, or K2 if data is from \citep{Szabo2020}; P: rotation period; $\delta$m: light curve amplitude; $H_V$: absolute magnitude; p$_V$: geometric albedo; D: estimated effective diameter; Redness: red (R) or less red (LR) indicators, based on \citet{Szabo2020} and data obtained from \citet{2021A&A...652A..59S}; $\overline{(g-r)}$: mean SDSS $(g-r)$ colour.   
        For (42237) we used the size and geometric albedo values from \citep{Grav2012} (marked by $^*$). In all other cases the population average value of p$_V$\,=\,0.061 \citep{Grav2012} is used to obtain the effective diameter.}
\label{table:targets}
\end{table*}

We also tested the identified periods using data from the Zwicky Transient Facility \citep{ZTF}. 
The tested periodicity is clearly evident in the cases of (42237) and (91273), confirming the period analysis from the TESS data. In addition, the overall shapes of the light curves are also similar. These consistencies strongly support the results for these two fast-rotating Hildas. Even in the case of the asteroid (237321), despite the significantly lower number of data points, a probability density analysis shows that the ZTF period is equal to the TESS period within the uncertainties.

   \begin{figure}
   \centering
    \includegraphics[width=\columnwidth]{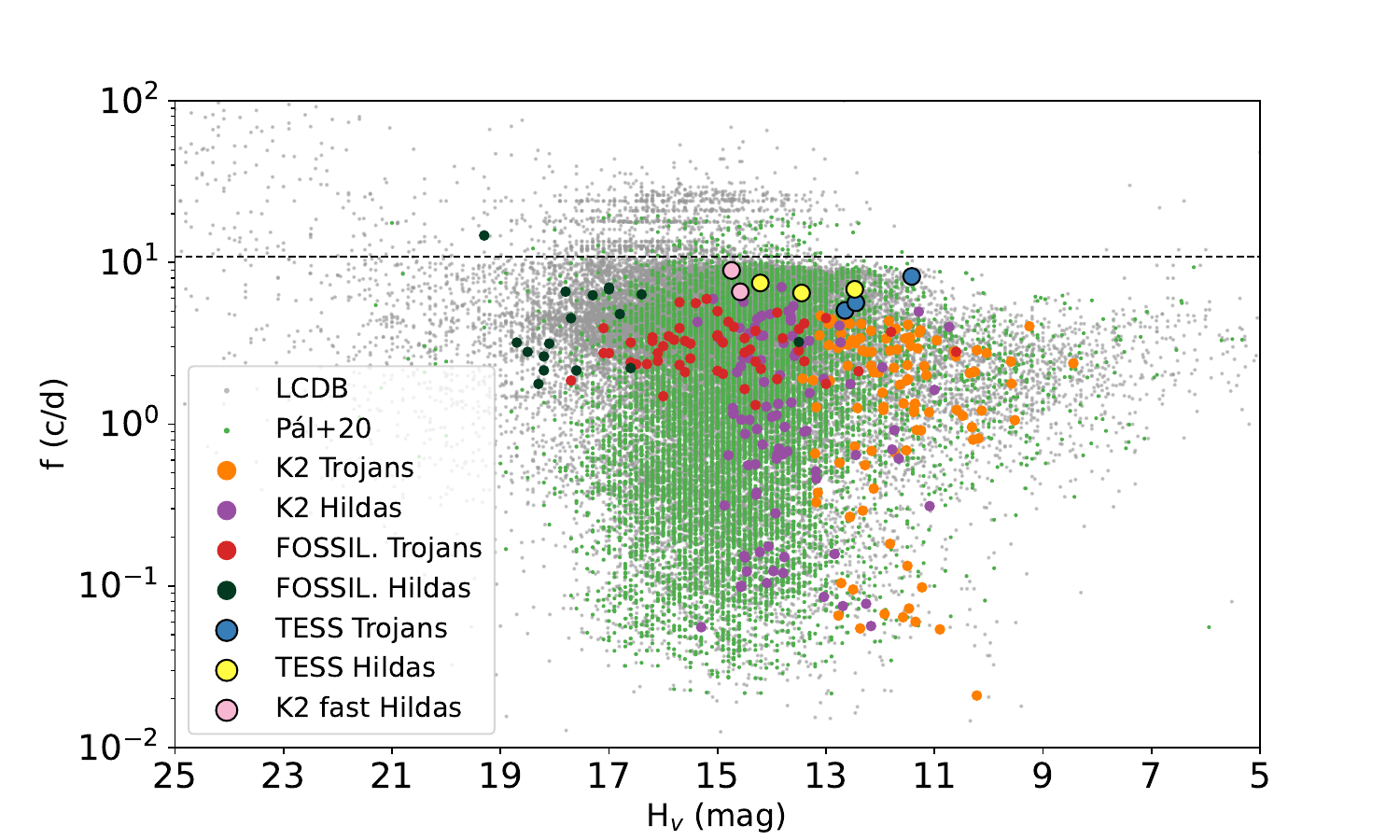}
    \includegraphics[width=\columnwidth]{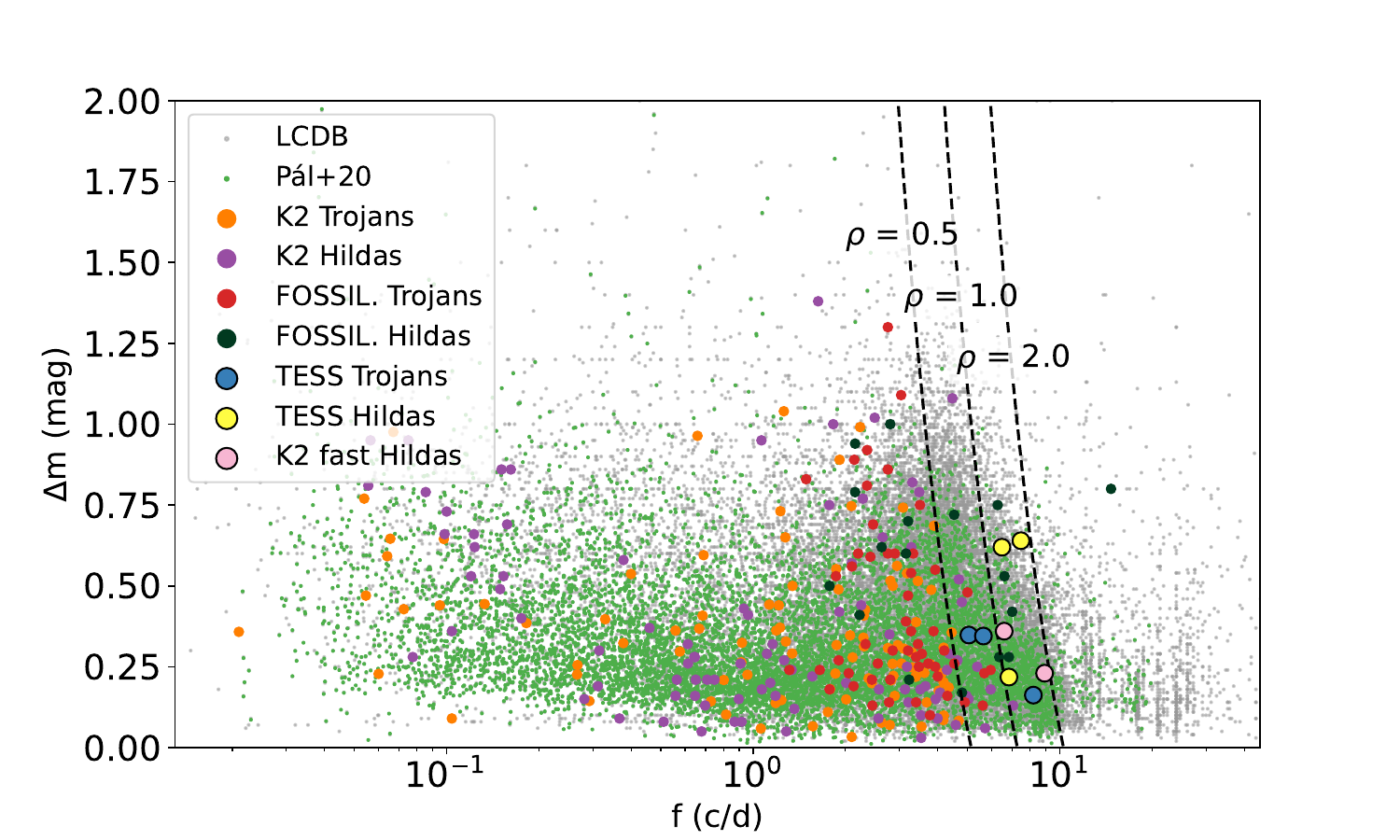}
   \caption{Rotational properties of inner Solar System asteroids. The top panel shows rotational frequency versus absolute magnitude, with the dashed horizontal line marking the rotational breakup limit (a 2.2-hour period) from \citet{Pravec2000}. The bottom panel plots light curve amplitude against rotational frequency, with dashed curves indicating constant critical densities from \citet{Pravec2000}. In both panels, different colors represent data from various sources, as shown in the legend. The three asteroids studied in this paper are highlighted with large yellow symbols, and the two fast rotators from K2 are marked with pink dots. The large blue dots indicate the fast-rotating Trojans from \citet{2025A&A...694L..17K}.
  }
              \label{fig:freqdistr}%
    \end{figure}

\section{Discussion and conclusions\label{sect:discussion}}

From the rotation period and light curve amplitude, a lower limit can be estimated for the bulk density, assuming a strengthless rubble-pile structure \citep{Pravec2000}. 
For our targets, the obtained rotation periods and the derived densities are presented in Fig.~\ref{fig:freqdistr}, together with data from large databases, including Jovian trojans and Hildas from the K2 mission \citep{Szabo2017,Szabo2020,Kalup2021}, and the FOSSIL survey \citep{Chang2021,Chang2022}. The resulting critical densities are between $\sim$1.1--1.9\,\gcc based on the \citet{Pravec2000} relation, using the periods and amplitudes listed in Table~\ref{table:targets}.

Similarly, as in \cite{2025A&A...694L..17K}, we may use a simple granular material model to replace the strengthless rubble-pile approximation. In this model, the cohesion necessary to keep the rotating body together is obtained from the criterion $\sqrt{J_2}\,\leq\,k-3sp$, where $J_2$ is the second invariant of deviator stresses, $k$ is the cohesion (shear stress at zero pressure), $s$ is the slope parameter and $p$ is the pressure.

\begin{figure}[ht!]
   \centering
    \includegraphics[width=\columnwidth]{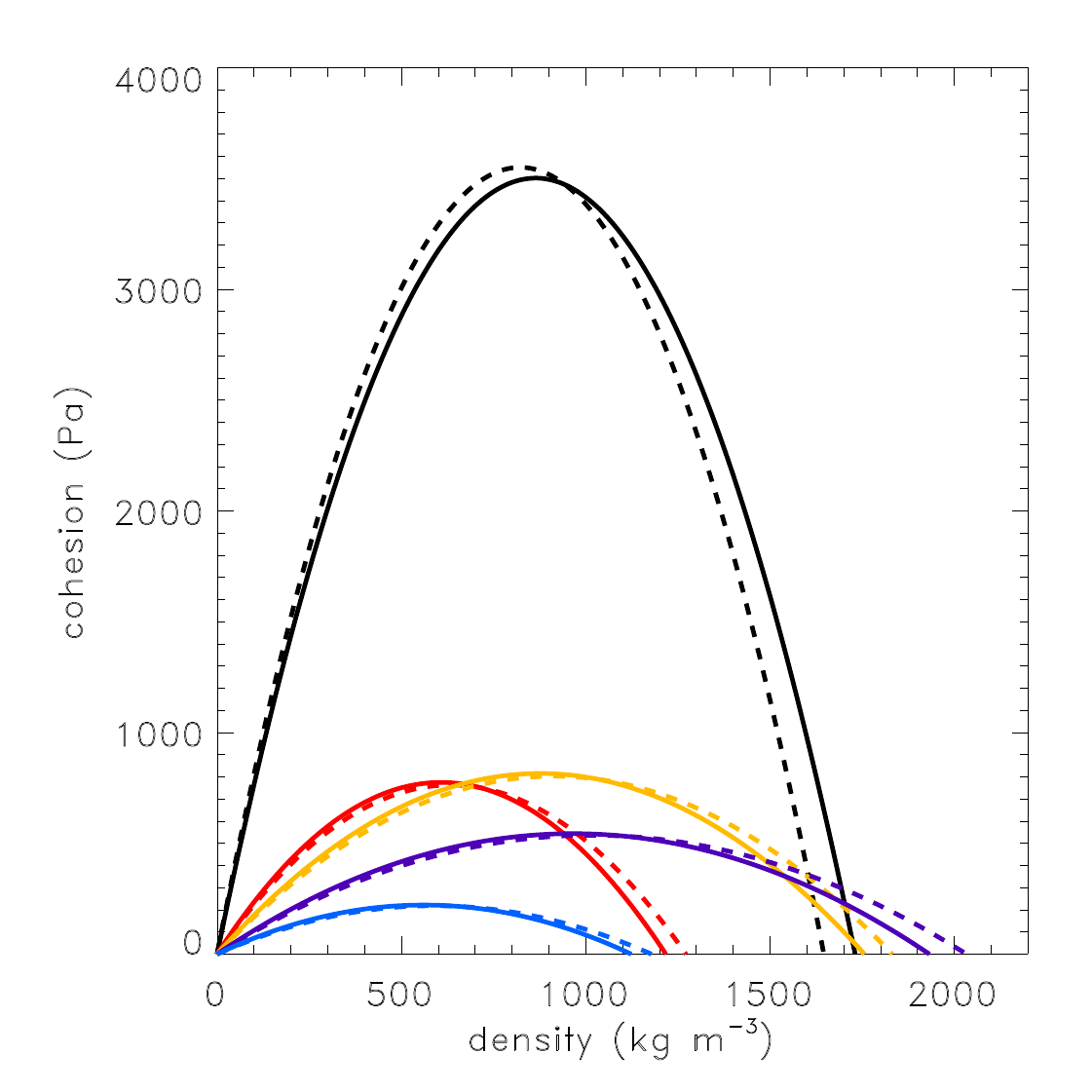}
   \caption{Cohesion versus density curves for our targets obtained using the Drucker-Prager criterion of failure. The colours correspond to the following targets: black -- (42237); red -- (91273); orange -- (237321); purple -- (207644); blue -- (208290). Dashed and solid curves correspond to friction angles of $\phi$\,=\,40\degr\,and 45\degr, respectively. 
   }
              \label{fig:stress}%
    \end{figure}

We approximate the shape of the asteroid with a triaxial ellipsoid with semi-axes $a\,>\,b\,\geq\,c$, where the ratio of the $a$ and $b$ semi-axes is simply obtained from the observed light curve amplitude, assuming that we see the system equator-on. We also note that this assumption provides an upper limit on the cohesion, as a larger maximum light curve amplitude, or a smaller b/a ratio would lead to a lower cohesion value at a specific density. The rotational breakup limit can be derived by using the Drucker-Prager criterion for failure \citep{Holsapple2004,Holsapple2007}. Following \cite{Holsapple2007}, we assume that the $b$ and $c$ semi-axes have equal lengths. For (42237), the absolute magnitude, visible range geometric albedo and size are obtained from \citep{Grav2012}. For the other targets, the effective diameters are obtained from the $H_V$ absolute magnitudes assuming a geometric albedo of $p_v$\,=\,0.061, a typical value among Hildas \citep{Grav2012} (see also Table~\ref{table:targets}). We assume an angle of friction of $\phi$\,=\,45\degr\, which corresponds to a slope parameter of $s$\,=\,0.356 \citep{Holsapple2007}, {as well $\phi$\,=\,40\degr\ ($s$\,=\,0.315) used in \citet{Polishook2016}, representative of the Lunar regolith \citep{Mitchell1974}.}
 The results are presented in Fig.~\ref{fig:stress}.

The densities associated with zero cohesion (k\,=\,0) are similar to those obtained from the strengthless rubble-pile case. For (42237), considerable cohesion is needed for densities below $\sim$1.5\,\gcc, in the order of 1--3\,kPa. The lunar regolith has cohesion typically between
100--1000\,Pa \citep{Mitchell1974}, and main
belt and near-Earth asteroids typically require cohesion on the order of a few hundred Pa \citep[see e.g.][]{Polishook2016}. However, (42237) is the largest asteroid (a size of $\sim$19\,km) in the sample. For the other, smaller targets (6--11\,km) the necessary cohesion remains below 1\,kPa for the entire range of densities ($\rho$\,$\lesssim$2\,\gcc). 
If we restrict the possible cohesion to a few hundred Pa, then these asteroids require densities in the range of 1.5--2.0\,\gcc.

The widely accepted scenario for the origin of the current Hilda population is related to the `giant planet instability' in the early solar system, in which a population of Hildas that existed prior to the instability event were destabilized \citep{Broz2011,2015AJ....150..186R}, and a new Hilda population was captured to the 3:2 resonance with Jupiter from a massive trans-Neptunian planetesimal disk \citep{2009Natur.460..364L,2016AJ....152...39V}. A small, $\lesssim$5\% fraction of these objects could have originated in the outer main belt region \citep{2015AJ....150..186R}, today dominated by C-type asteroids \citep{DeMeo2014}.  

Among Jovian trojans, (3548)~Eurybates is a large C-type asteroid, well within the L4 stability region. Most known C-type Jovian trojans are also members of the Eurybates asteroid family \citep{Broz2011}, which comprises only a small fraction of Jovian trojans.
Based on Sloan Digital Sky Survey colours, \citet{2008Icar..193..567G} obtained that $\sim$16\% of the Hildas investigated belong to the `broad C-class', a significantly higher fraction than that of the C-type asteroids among Jovian trojans. C-type asteroids have a notably lower density limit of $\sim$1.33\,\gcc\, than that of other asteroid taxonomy classes, except the D-class \citep[$\sim$1\,\gcc,][]{2013Icar..226..723D,2015ApJS..219...27C}, the common type among Hildas. Large C-type main belt asteroids have typically even larger densities of $\sim$1.6\,\gcc\ \citep{Vernazza2021}. We identified five fast-rotating (P\,$\leq$4\,h) Hildas in the K2 and TESS samples, a significantly larger number than a single Jovian trojan \citep{2025A&A...694L..17K} in this rotation period range. The higher occurrence rate of fast-rotating Hildas may be explained by a larger fraction of C-type asteroids. 

If the spin frequency distribution of a population can be described by a Maxwellian distribution assuming collisional relaxation \citep{Pravec2000,Szabo2020}, we can calculate the expected fraction of fast rotators (in our case P\,$\leq$\,4.0\,h). 
In a simple model, we consider two populations: a population of C-type asteroids including fast rotators (allowed by their larger densities), and another population which completely lacks fast rotators due to their low densities. We further assume that the common frequency distribution can be described by combining the Maxwellian distributions of these two populations, that the common peak frequency of this combined distribution is f\,=\,2.5\,c/d \citep{Szabo2020}, and we allow the fraction of the C-type asteroids and the related mode (peak) frequency to vary. Using the entire K2 and TESS sample, we estimate that the occurrence rate of fast rotators is $\sim$1\%.

\begin{figure}[ht!]
    \centering
    \includegraphics[width=\columnwidth]{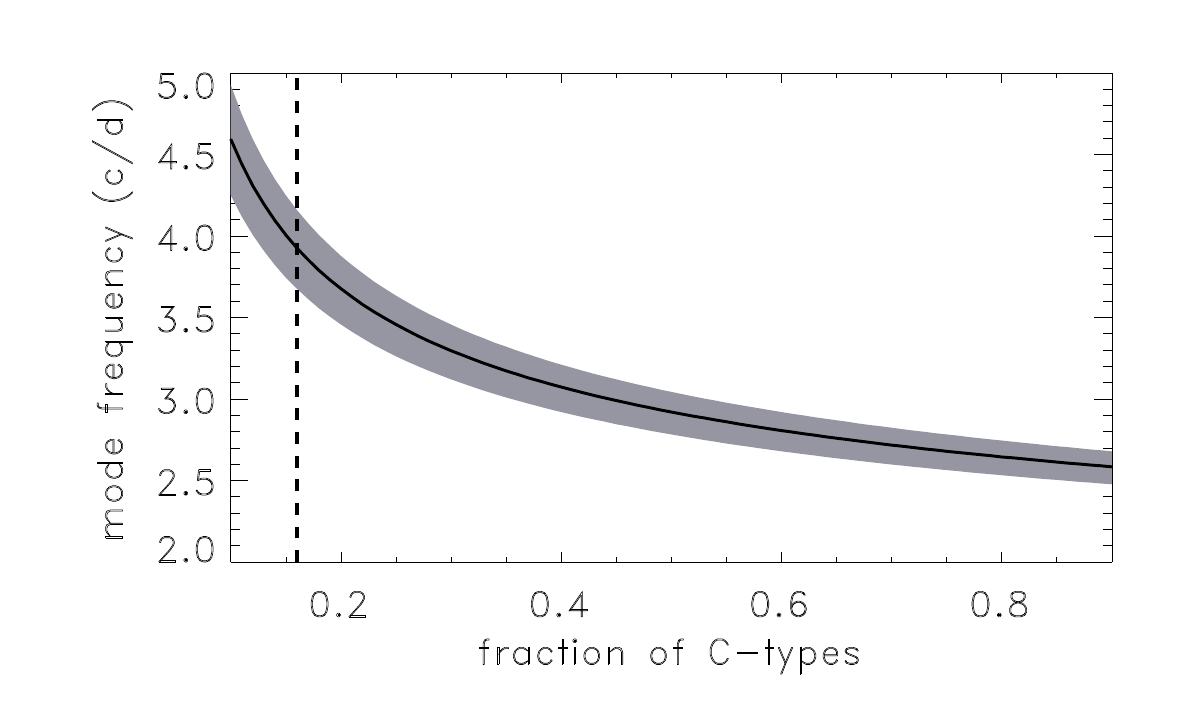}
    \includegraphics[width=\columnwidth]{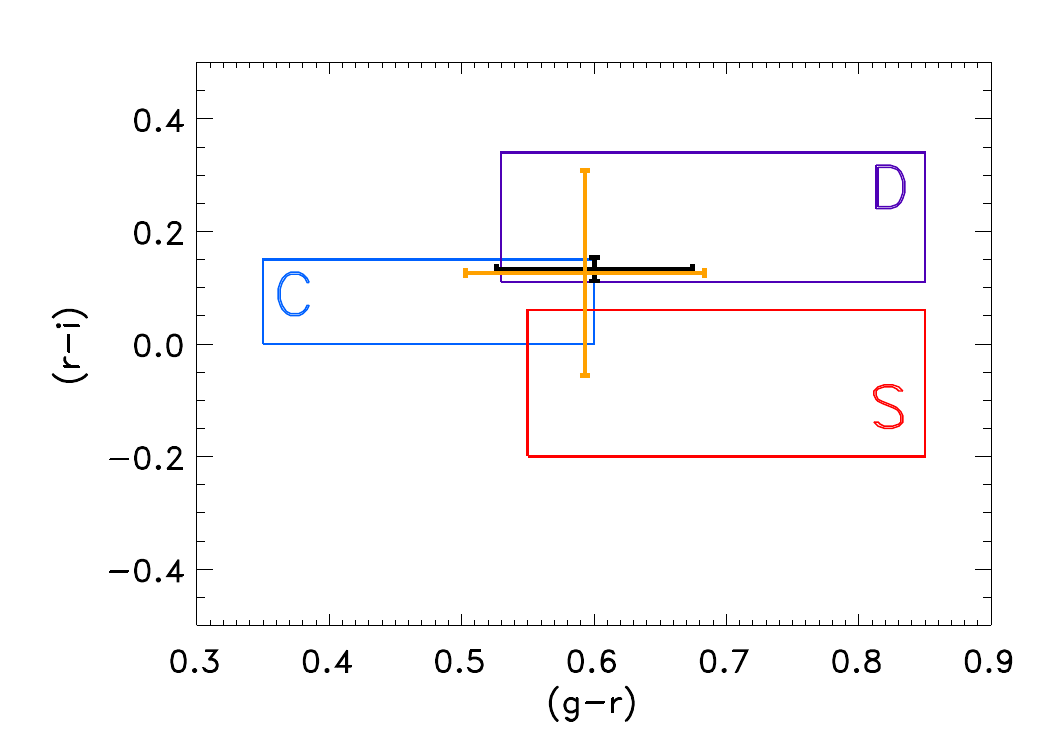}
    \caption{Top panel: Maxwellian mode frequency of the suspected C-type population as a function of the fraction of this population among Hildas. The solid curve corresponds to a fast-rotator (P\,$\leq$4\,h) fraction of 1\% in the whole population. The gray-shaded region marks the region corresponding to the 0.8-1.2\% C-type fraction curves. The vertical dashed line marks the C-type fraction obtained by \citet{2008Icar..193..567G}.
    Bottom panel: SDSS $(g--r)$ and $(r--i)$ colors and their uncertainties for (42237) (black cross) and (207644) (orange cross). The rectangles with labels mark the approximate areas of the main taxonomy types C, D and S (blue, purple and red, respectively), as defined by \citet{2021A&A...652A..59S}.}
    \label{fig:maxw}
\end{figure}
We calculated the mode frequencies of the C-type subpopulation as a function of the C-type fraction (Fig.~\ref{fig:maxw}).
Assuming a realistic C-type fraction of 10-16\%, the mode frequencies of the C-type population are in the range of $\sim$4-5\,c/d, which corresponds to rotational frequencies of 4.8-6.0\,h, using a range of fast rotator fractions of 0.8-1.2\%. These frequencies are
comparable with the mode frequency observed in the main belt \citep[f\,$\approx$\,4\,c/d,][]{Pal2020}, i.e., the existence of such a population of C-type asteroids within the Hilda group may explain the observed rate of fast rotators among Hildas.  

Current broad-band photometric measurements cannot strongly constrain the possible taxonomy of the fast-rotating Hildas discussed in this paper. Combined ZTF photometry (Sect.~\ref{sect:obs}), Pan-STARRS 
data\footnote{https://ps1images.stsci.edu/ps1-moving-database-CADC.html} 
when (nearly) parallel observations in multiple bands were available, 
and data from the moving object archive of the Sloan Digital Sky Survey \citep{2021A&A...652A..59S} provide SDSS colors of $(g-r)$\,=\,0.60$\pm$0.07 and $(r-i)$\,=\,0.13$\pm$0.02 for (42237), placing it at the boundary of the C and D-type asteroids in the respective color-color plane, as presented in Fig.~\ref{fig:maxw} \citep[see][for the definition of asteroid taxonomic types in SDSS colours]{Ivezic2001,2021A&A...652A..59S}. 
Similarly, (207644) is also at the edge of the C-type boundary ($(g-r)$\,=\,0.59$\pm$0.09 and $(r-i)$\,=\,0.13$\pm$0.18), however, these colors also have large uncertainties. For the other three targets only $(g-r)$ colors could be derived from ZTF photometry (see Table~\ref{table:targets}), and in all cases $(g-r)$\,$<$\,0.6, i.e., they could potentially be C-type asteroids. We also note that the large light curve amplitudes of these asteroids contribute significantly to the color uncertainty, as survey measurements in the different bands are not performed in parallel. Dedicated photometric and spectroscopic measurements are needed to obtain the taxonomy of these fast-rotating Hildas. Such observations can possibly reveal their origin and further constrain the processes that created the currently observable Hilda population in the early solar system.

\begin{acknowledgements}
Acknowledgements~
This paper includes data collected by the TESS mission. Funding for the TESS mission is provided by the NASA's Science Mission Directorate. The research leading to these results has received funding from the K-138962, SNN-147362, KKP-137523 and TKP2021-NKTA-64 grants of the National Research, Development and Innovation Office (NKFIH, Hungary). This work made use of Astropy (\url{http://www.astropy.org}), a community-developed core Python package and an ecosystem of tools and resources for astronomy \citep{astropy:2013, astropy:2018, astropy:2022}. This work used GNU Parallel \citep{tange2022}. This research made use of NASA’s Astrophysics Data System Bibliographic Services. 
We thank the hospitality of F\H{o}nix Ba\-da\-csony where this project was carried out.
\end{acknowledgements}

\setlength{\bibsep}{1pt}

\bibliography{refs}
\bibliographystyle{aa}

\label{LastPage}
\end{document}